\documentclass[
]{ceurart}

\usepackage{listings}
\usepackage{ragged2e}
\usepackage{blindtext}
\usepackage{hyperref}
\usepackage{amsmath,amssymb,amsfonts}
\usepackage{algorithmic}
\usepackage{multirow}
\usepackage{graphicx}
\usepackage{textcomp}
\usepackage{xcolor}
\usepackage{enumitem}

\begin{document}

\copyrightyear{2024}
\copyrightclause{Copyright for this paper by its authors.
  Use permitted under Creative Commons License Attribution 4.0
  International (CC BY 4.0).}

\conference{CAMLIS'24: Conference on Applied Machine Learning for Information Security,
  October 24--25, 2024, Arlington, VA}

\title{Embedding-based classifiers can detect prompt injection attacks}

\author[1]{Md. Ahsan Ayub}[
orcid=0000-0002-1345-0110,
email=ahsan.ayub@vumc.org
]
\address[1]{Enterprise Cybersecurity, Vanderbilt University Medical Center, Nashville, TN, USA}

\author[2]{Subhabrata Majumdar}[
orcid=0000-0003-3529-7820,
email=subho@vijil.ai,
]
\address[2]{Vijil, Seattle, WA, USA}

\begin{abstract}
Large Language Models (LLMs) are seeing significant adoption in every type of organization due to their exceptional generative capabilities. However, LLMs are found to be vulnerable to various adversarial attacks, particularly prompt injection attacks, which trick them into producing harmful or inappropriate content. Adversaries execute such attacks by crafting malicious prompts to deceive the LLMs. In this paper, we propose a novel approach based on embedding-based Machine Learning (ML) classifiers to protect LLM-based applications against this severe threat. We leverage three commonly used embedding models to generate embeddings of malicious and benign prompts and utilize ML classifiers to predict whether an input prompt is malicious. Out of several traditional ML methods, we achieve the best performance with classifiers built using Random Forest and XGBoost. Our classifiers outperform state-of-the-art prompt injection classifiers available in open-source implementations, which use encoder-only neural networks. \\
\textcolor{red}{Warning: This paper discusses and contains language that could be considered inappropriate for readers.}
\end{abstract}


\maketitle

\section{Introduction}
Large Language Models (LLMs) have been widely adopted to streamline daily tasks that need automation, such as text (including code) generation \cite{tang2023science, liu2024your}, text summarization \cite{chang2023booookscore}, sentiment analysis \cite{zhang2023enhancing}, AI chatbots \cite{kim2023chatgpt}, and machine translation \cite{he2024exploring}. 
Compared to traditional software, LLM, generative AI (genAI) applications, and agents embody a much broader attack surface that adversaries can exploit for malicious purposes. For example, LLMs are found to produce contents containing gender and racial biases, toxicity, disinformation, and misinformation \cite{zhuo2023exploring, liang2022holistic, bai2022training}. To further explain, the models are designed to generate response based on the supplied prompts. By crafting malicious prompts, attackers attempt to override LLM developers' instructions to exploit the baseline model.

In this paper, we focus on examining prompts that lead to successful prompt injection attacks. Specifically, we investigate the behavior of malicious prompts that attempt prompt injection versus benign prompts in the embedding space. With the goal of developing effective embedding-based approaches to safeguard genAI applications from prompt injection attacks, we ask the following important research questions:

{\textit{\textbf{RQ1.}}} Are there any dissimilarities between benign and malicious prompts?

{\textit{\textbf{RQ2.}}} Can we effectively identify malicious prompts to thwart prompt injection attacks?

To address the questions above, we curate a dataset of 467,057 malicious and benign prompts. We obtain their embeddings based on three state-of-the-art embedding models and use this data to gain insights into the overall behavior of malicious prompts.

The major contributions of our paper are as follows:
\begin{itemize}[nolistsep]
    \item We investigate the distributional differences of benign and malicious embeddings generated using three embedding models: from the API-only OpenAI \texttt{text-embedding-3-small}, and the open-source models \texttt{gte-large}, and \texttt{all-MiniLM-L6-v2}.
    \item Using the embeddings as input datasets, we build a suite of supervised machine learning (ML) classifiers to detect prompt injection attacks.
    \item Across several metrics, we compare the performance of embedding-based classifiers' with state-of-the-art deep learning based prompt injection classifiers.
\end{itemize}

\noindent Our implementation, along with the curated datasets used for evaluation, is available on GitHub\footnote{\url{https://github.com/AhsanAyub/malicious-prompt-detection}}.

The rest of the paper is organized as follows: Section 2 discusses the background of our research. The discussion of experimental methodology, including the construction of the dataset, and our empirical findings are described in Sections 3 and 4, respectively. We share the related work in this field in Section 5 and list the limitations of our study, as well as future work, in Section 6. Finally, we provide the conclusion of this work in Section 7.

\section{Background and Related Work}
This section describes some background and related work, as context for the subsequent sections.

\subsection{Prompt Injection Attacks}
SQL injections and Cross-Site Scripting (XSS) attacks are among the most commonly found cyber threats, where attackers craft payloads to disrupt the routine execution of a program \cite{boyd2004sqlrand, gupta2017cross}. With the proliferation of genAI, adversaries can carry out similar attacks by injecting LLMs with malicious prompts. In genAI applications, users can utilize the extensible functionalities of LLMs via natural language-based prompts to generate desired outputs. Attackers exploit this interaction pattern by supplying crafted prompts to cause LLMs to perform undesired actions \cite{perez2022ignore, yu2023assessing}. These malicious prompts can be supplied as inputs either by as malicious users, or by attackers modifying benign user-provided prompts through man-in-the-middle attacks. A successful prompt injection attack leads to unintended consequences, such as the exposure of underlying system prompts, disclosure of private data, and attackers gaining unauthorized access to functionalities the LLM is authorized to perform but the user is not \cite{liu2023prompt}.

Another example of such attacks is when an LLM takes input from external sources, such as websites or files---adversaries inject malicious prompts to hijack the context. This is known as an indirect prompt injection attack. The goal of this attack is to extract sensitive, harmful, or unwanted information from the LLMs \cite{greshake2023not}. Real attackers do not need to possess deep technical knowledge about how the model is built, or compute gradients, to trick the LLM application into responding to a distinct set of queries with the intent of compromising it \cite{apruzzese2023real}.

\subsection{Embedding Models}
\begin{figure}[t]
    \centering
    \includegraphics[width=0.75\columnwidth]{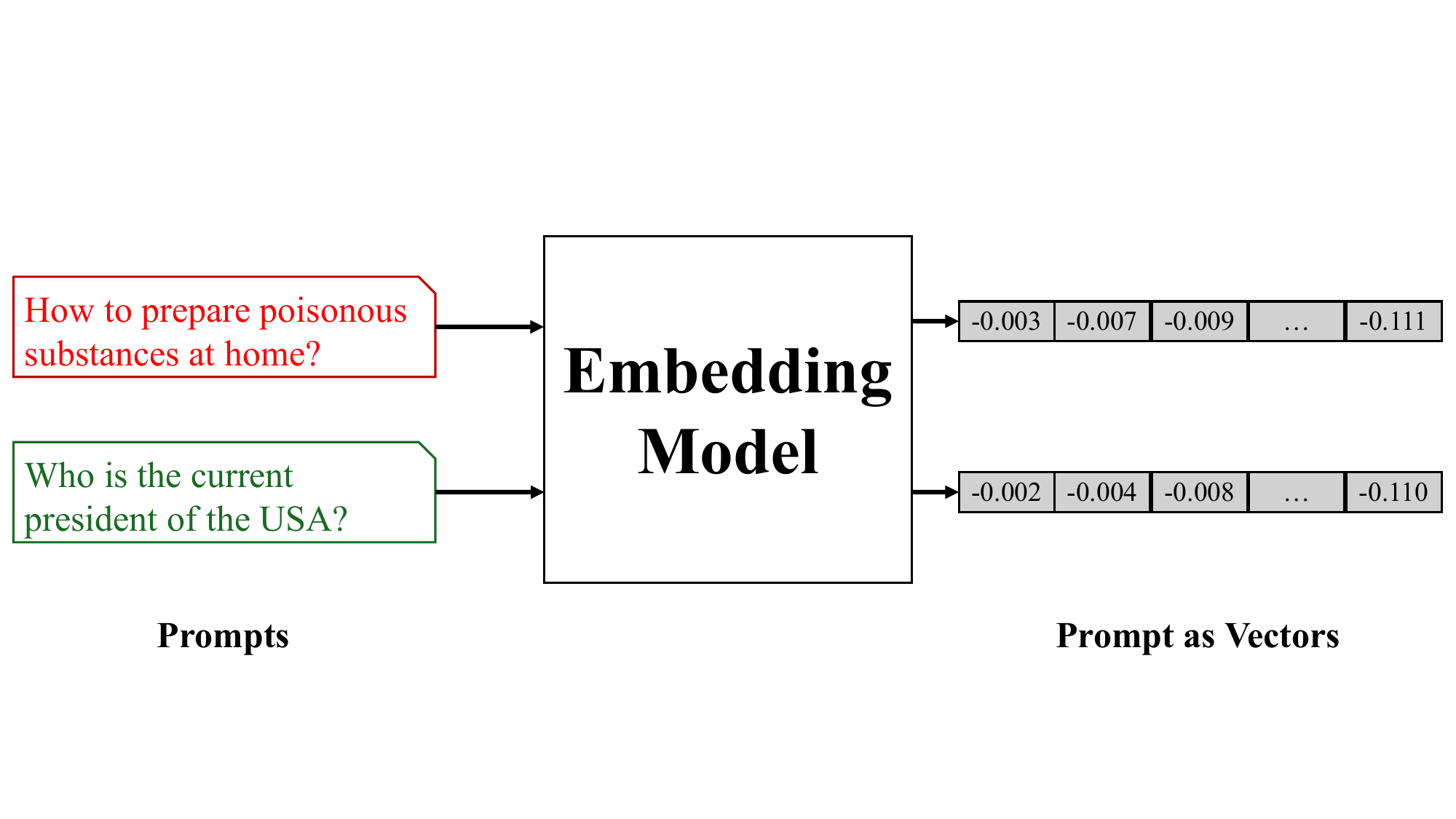}
    \caption{A schematic diagram of prompts and their embeddings.}
    \label{fig:embedding_generation}
\end{figure}

Prompts are primarily constructed using natural language, and their size can vary widely. We apply embedding models to convert the textual data in prompts into dense representations in a multi-dimensional space of fixed dimension \cite{neelakantan2022text}. To explain further, each prompt is transformed into a fixed-length sequence of floating-point numbers. Such numerical representations enable us to create a vector database derived from a list of prompts. To accomplish these tasks, we select the following embedding models: \texttt{text-embedding-3-small} from OpenAI\footnote{\url{https://platform.openai.com/docs/guides/embeddings}}, and the open-source models \texttt{gte-large} hosted on OctoAI\footnote{\url{https://octo.ai/blog/introducing-octoais-embedding-api-to-power-your-rag-needs}}, as well as the well-known \texttt{all-MiniLM-L6-v2}. For brevity we refer to them as OpenAI, GTE, MiniLM from here on. It is important to note that all of these embedding models are contextual representation models  \cite{liu2019linguistic}. This means that each word is placed in the vector space based on the input context. For example, the word ``apple'' has a static meaning of being a fruit. However, given a specific input context, it can also refer to a technology company. Our selected embedding models are equipped to capture the context of each word in a prompt\cite{asudani2023impact}. We illustrate an example in Fig. \ref{fig:embedding_generation}.

The length of the embedding vector for OpenAI is 1536, which means that any size of textual data is mapped to a 1536-dimensional dense numerical vector space. Additionally, the embedding vector sizes for GTE and MiniLM are 1024 and 384, respectively.

\subsection{Related Work}
In recent past, genAI threats---especially prompt injection attacks---have received a significant attention from AI security practitioners and researchers. A successful attack enables adversaries to override intended use guidelines of an LLM application to generate violating content along different directions, such as hate speech and discrimination, profanity, sexual, violent, and unsafe content, controversial topics, illegal activities, self-harm, harassment, and unethical actions\footnote{\url{https://www.robustintelligence.com/ai-security-and-safety-taxonomy}}. AI security researchers have come up with numerous techniques that adversaries can utilize to perform prompt injection attacks \cite{kang2023exploiting, shen2023anything, wei2024jailbroken, samvelyan2024rainbow, perez2022ignore, bai2024special}. Therefore, it is important to defend genAI applications against such attacks.

We break down prior research on prompt injection detection into two categories.

\paragraph{Guardrail-based (AI Firewall) Defense.}
\citet{alon2023detecting} used a perplexity-based approach to detect malicious prompts by computing perplexity to estimate text quality. The injection of instructions or data into prompts influences quality and results in a high perplexity value. \citet{jain2023baseline} divided textual data into contiguous windows for perplexity calculation to check whether any window's perplexity exceeds the threshold. \citet{chen2024struq} examined how separating prompts and supplied data enables LLMs to become more robust against prompt injection. \citet{yi2023benchmarking} found that placing a special delimiter between the prompt and data allows LLMs to distinguish between malicious external content and user instructions, thereby preventing harmful outputs. \citet{schulhoff2023ignore} discovered that adding extra text to the prompt to make LLMs aware of prompt injection attacks would also be effective.

\paragraph{LLM-based Defense.}
Recent LLMs, such as GPT-4o and the Gemini and Claude families of models, show a propensity of rejecting harmful prompts incorporated through safety training \cite{achiam2023gpt,ganguli2022red,wei2024jailbroken}. LLMs may also be used as detectors designed to identify malicious prompts through their training \cite{belrose2023eliciting}. AI security researchers have showed that it is possible to detect prompt injection by providing explicit instructions to LLMs, such as \textit{``..Your job is to analyze whether the input prompt is safe...''} and using this model as an LLM-as-judge to evaluate input prompts \cite{armstrong5using}. Finally, traditional encoder-only NLP models, such as the ones using a DeBERTa arcchitecture that utilizes disentangled attention and an enhanced mask decoder, can detect prompt injection and jailbreak attacks \cite{deberta-v3-base-prompt-injection-v2,promptguard}.

\paragraph{Our Approach.}
To the best of our knowledge, our study is the first attempt to investigate the effectiveness of embedding-based classifiers in detecting malicious prompts. Although a lot of work has already been published in this area, we did not find any research on embeddings of malicious prompts and their efficacy in leading to successful detection. We hope that this research will make singnificant contribution to the AI safety and security domain by extending and reproducing our experiments.

\section{Methodology}
\subsection{Dataset Construction}
The dataset used in our experiments is curated from open-source datasets containing malicious and benign prompts pertaining to prompt injection attacks (Table \ref{tab:dataset_details}). 
In total, we acquire a total of 553,185 numbers of malicious and benign prompts. After deduplication, we end up with a total of 467,057 unique prompts, of which 109,934 (23.54\%) are malicious. Each prompt is assigned a unique identifier and a source to indicate its origin. Therefore, the dataset columns appear as follows: ID, Source, Text, and Label (0 to denote benign, 1 for malicious). Using the \texttt{train\_text\_split} method\footnote{\url{https://huggingface.co/docs/datasets/v1.8.0/processing.html\#splitting-the-dataset-in-train-and-test-split-train-test-split}}, we split this dataset into 80\% training and 20\% test sets. To ensure equal proportion of the malicious and benign labels across splits, we use stratified sampling. 

\begin{table*}[t]
    \centering
    \caption{Hugging Face datasets used in our study.}
    \begin{tabular}{l c}
        \hline
        Dataset (User: Title) & \# fo Prompts \\ \hline
        imoxto: Prompt Injection cleaned dataset & 535,105 \\
        reshabhs: SPML Chatbot Prompt Injection & 16,012 \\
        Harelix: Prompt Injection Mixed Techniques & 1,174 \\
        JasperLS: Prompt Injections & 662 \\
        fka: Awesome Chatgpt Prompts & 153 \\
        rubend18: ChatGPT Jailbreak Prompts & 79 \\ \hline
    \end{tabular}
    \label{tab:dataset_details}
\end{table*}

We develop a data pipeline using Python 3.11 to generate the embeddings for all prompts. With OpenAI's API key, we submit each prompt to get its embedding through \texttt{text-embedding-3-small} model. To obtain the GTE embeddings, we use the \texttt{thenlper/gte-large} model\footnote{\url{https://huggingface.co/thenlper/gte-large}}, accessed remotely through the serverless endpoint on OctoAI. For the MiniLM embeddings, we download the \texttt{sentence-transformers/all-MiniLM-L6-v2}\footnote{\url{https://huggingface.co/sentence-transformers/all-MiniLM-L6-v2}} model and host it locally. This approach allowed us to construct three separate tabular datasets composed of embeddings based on each of the embedding models.

Embeddings consist of fixed-length numerical representations. Therefore, we convert them into column values from lists. For instance, OpenAI generates an embedding vector consisting of 1,536 floating-point numbers for each prompt. We organize these numbers into 1,536 columns, treating each vector item as a separate column value. Consequently, the final embedding dataset generated by OpenAI comprises 1,539 features, with ID, Source, and Label as additional columns. Similarly, the embedding datasets for OctoAI and MiniLM consist of 1,027 and 387 features, respectively.

\subsection{Experimental Setup}

\paragraph{Methodology to Address RQ1: Visualization of Embeddings after Dimension Reduction.}
The embeddings provide us with high-dimensional tabular datasets. We apply Principal Component Analysis (PCA) \cite{abdi2010principal}, t-Distributed Stochastic Neighbor Embedding (t-SNE) \cite{van2008visualizing}, and Uniform Manifold Approximation and Projection (UMAP) \cite{mcinnes2018umap} to reduce these dense data distributions to a two-dimensional plane for visualization. This approach will help us investigate whether there are clear decision boundaries that can separate malicious prompts from benign ones.

We start our analysis with PCA, a linear dimensionality reduction approach using Singular Value Decomposition of the data \cite{stewart1993early}. We employ \texttt{sklearn} \cite{scikit-learn}, a Python machine learning package, to apply PCA and project the embedding-based columns into a 2-dimensional space.

Next, we use t-SNE to visualize our high-dimensional embedding distributions. It uses a nonlinear dimensionality reduction technique, unlike PCA. Similar to PCA, we use sklearn to execute the tasks. One of the major hypermeters of this algorithm is ``perplexity'', which is a guess about the number of close neighbors each point has \cite{wattenberg2016use}. Typically, its suggested values range between 5 and 50.

Finally, we examine the visualizations obtained using UMAP, which is based on manifold theory \cite{mac2013categories}. It seeks a low-dimensional representation of embeddings with an equivalent fuzzy topological structure. The algorithm operates in two phases: first, constructing a weighted k-nearest neighbor graph, and second, computing a low-dimensional layout of this graph. Variations among algorithms in this class lie in the specific methods used for graph construction and layout computation \cite{mcinnes2018umap}. We utilize its Python package to run our experiments\footnote{\url{https://umap-learn.readthedocs.io/en/latest/basic\_usage.html}}.

\paragraph{Methodology to Address RQ2: Binary Classification.}
To detect the malicious prompts, we train classifiers using three traditional ML methods: Logistic Regression \cite{schmidt2017minimizing}, eXtreme Gradient Boosting (XGBoost) \cite{chen2016xgboost}, and Random Forest \cite{breiman2001random}. We use sklearn to apply Logistic Regression and Random Forest. To implement XGBoost\footnote{\url{https://xgboost.readthedocs.io/en/stable/python/index.html}}, we use its corresponding Python package to run experiments. The goal of employing these classifiers is to train them on the train splits of each embedding dataset, encompassing both benign and malicious prompts, enabling the algorithms to discern underlying patterns. We evaluate the out-of-sample efficacy of each classifier on the test splits of the respective embedding datasets.

\section{Results}
In this section, we report the findings of our experiments.

\subsection{Answer to RQ1: Are benign and malicious prompts dissimilar in the embedding space?}
\begin{figure*}[t]
    \centering
    \includegraphics[width=.33\textwidth]{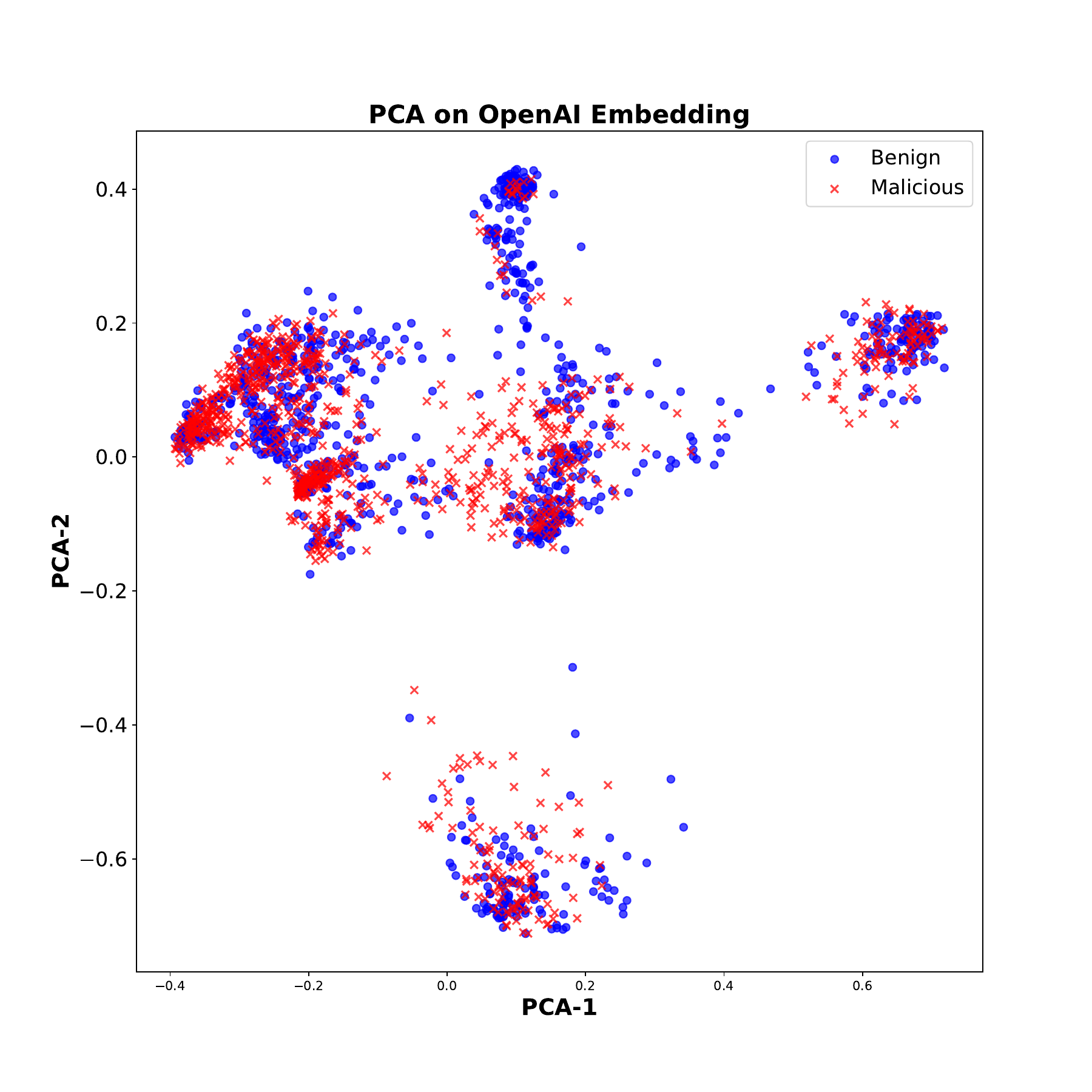}
    \includegraphics[width=.33\textwidth]{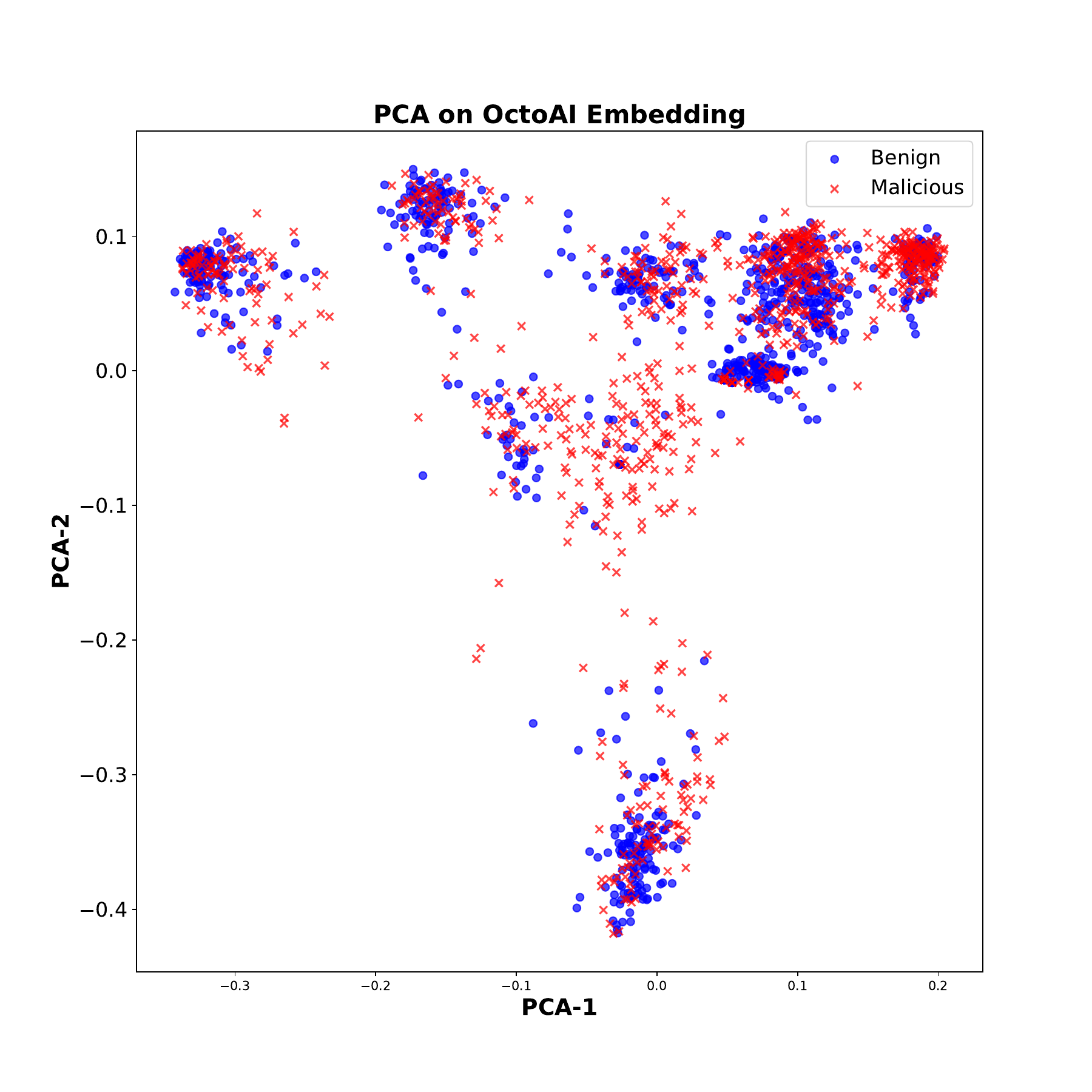}
    \includegraphics[width=.33\textwidth]{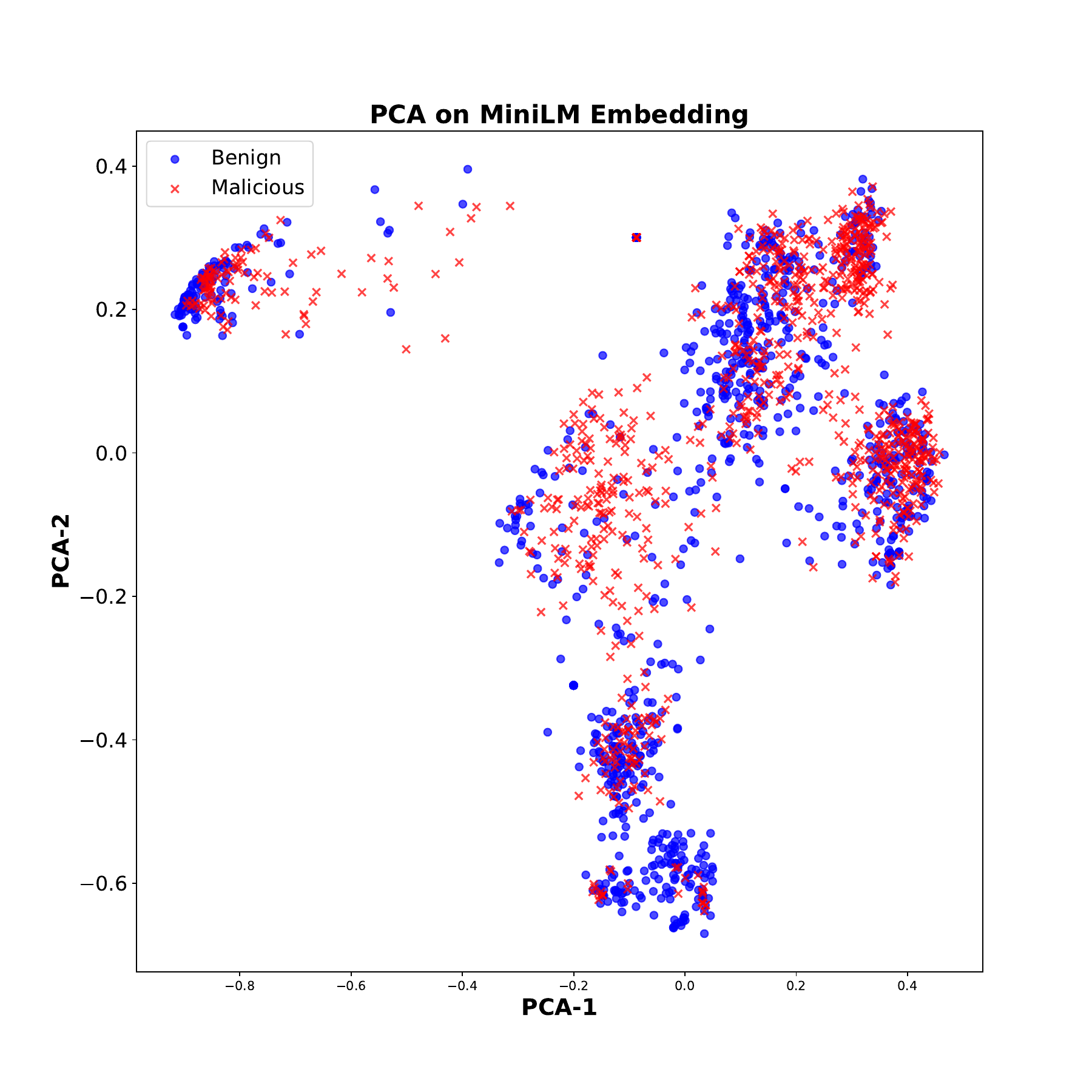}
    \caption{Visualization of OpenAI (Left), GTE (Middle), and MiniLM (Right) embedding distribution after applying Principal Components Analysis (PCA).}
    \label{fig:pca_on_embedding}
\end{figure*}

\begin{figure*}[t]
    \centering
    \includegraphics[width=.33\textwidth]{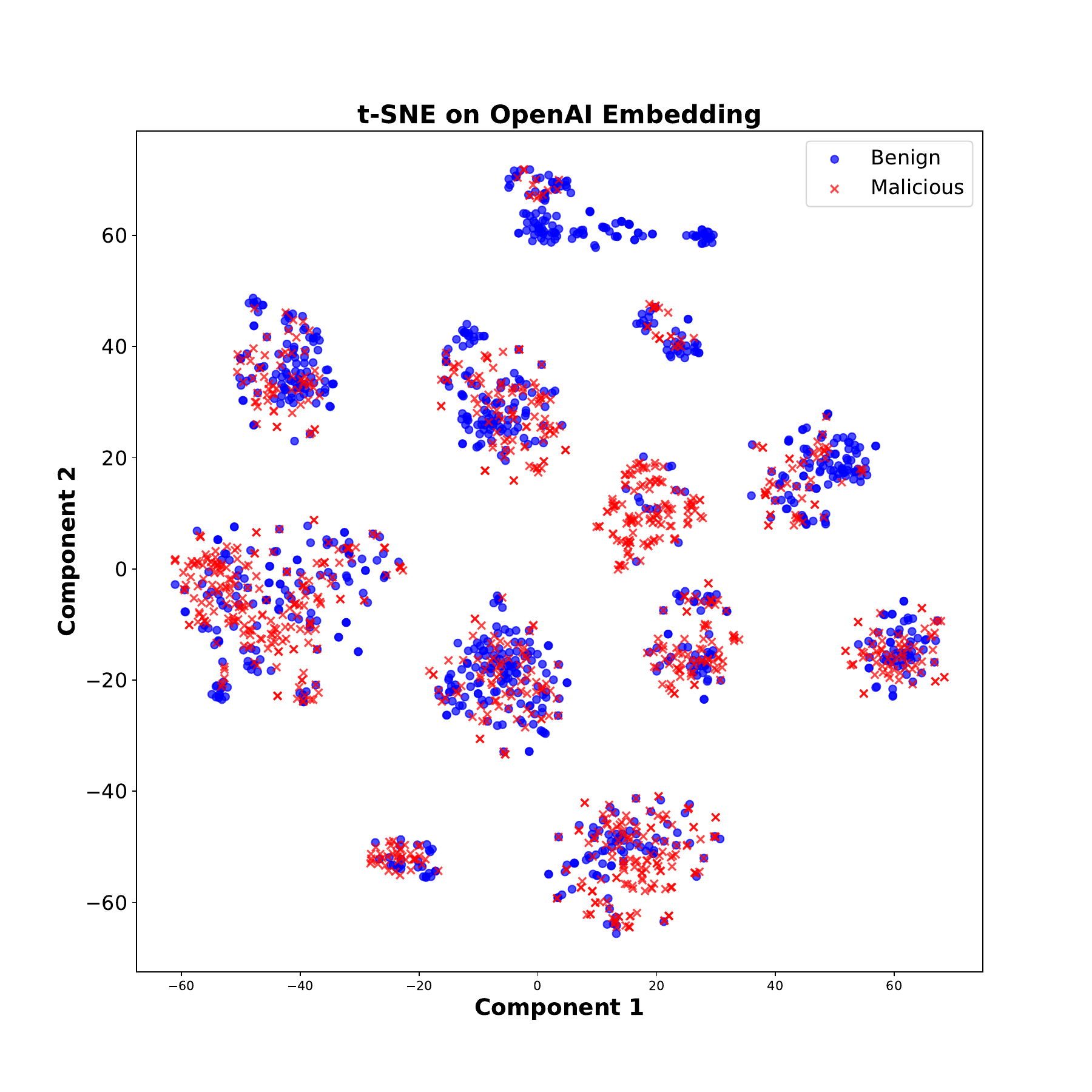}
    \includegraphics[width=.33\textwidth]{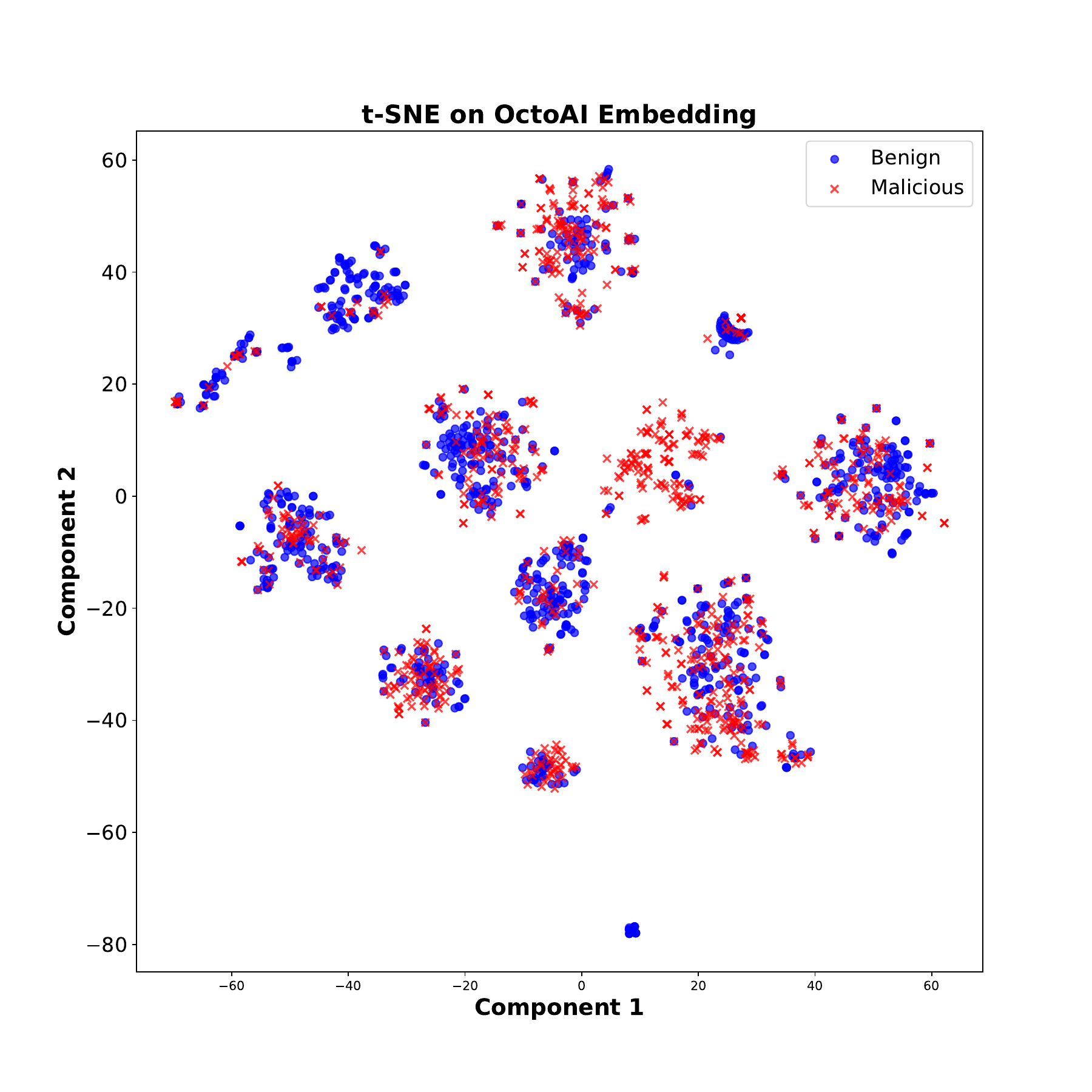}
    \includegraphics[width=.33\textwidth]{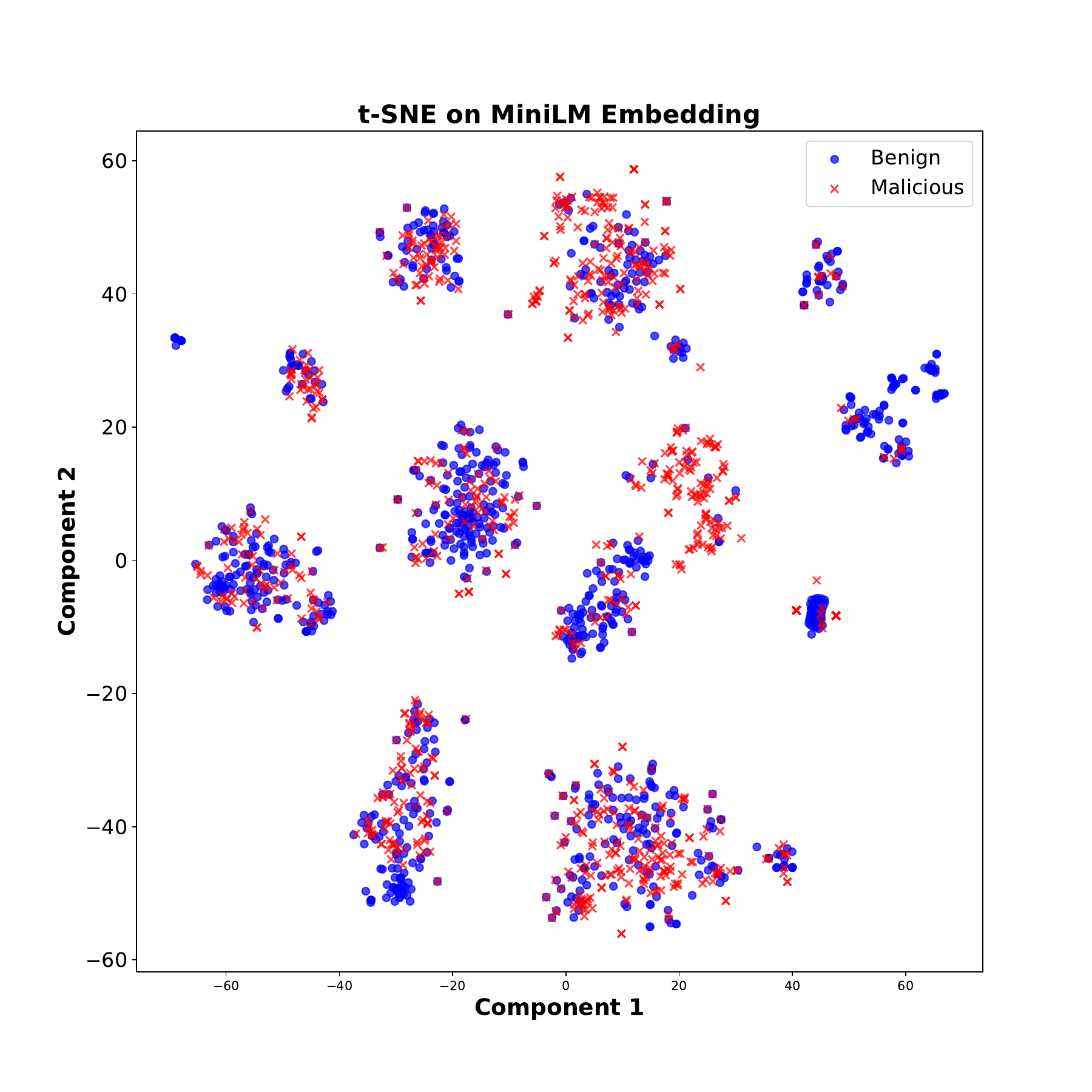}
    \caption{Visualization of OpenAI (Left), GTE (Middle), and MiniLM (Right) embedding distribution after applying T-distributed Stochastic Neighbor Embedding (t-SNE).}
    \label{fig:tsne_on_embedding}
\end{figure*}

\begin{figure*}[t]
    \centering
    \includegraphics[width=.33\textwidth]{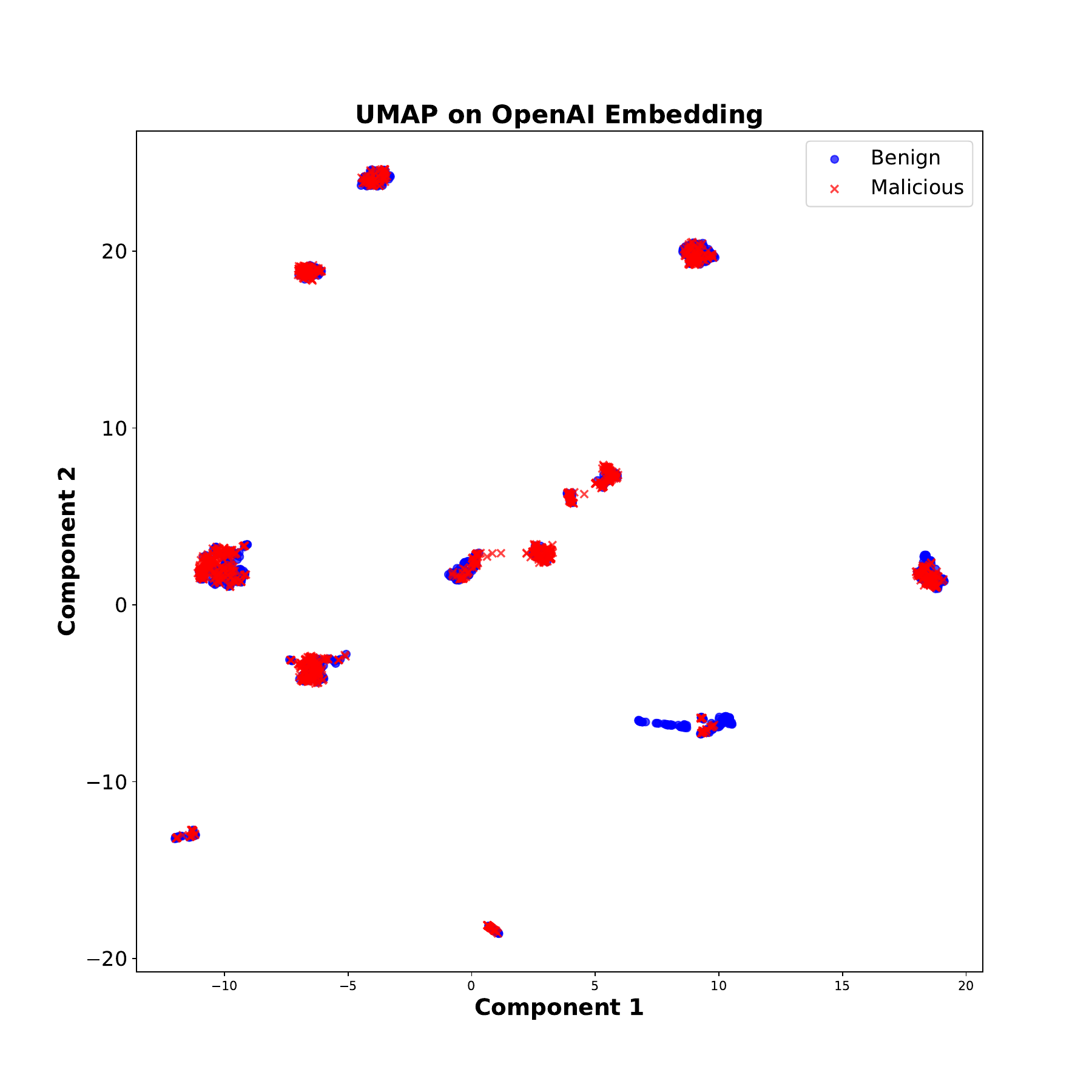}
    \includegraphics[width=.33\textwidth]{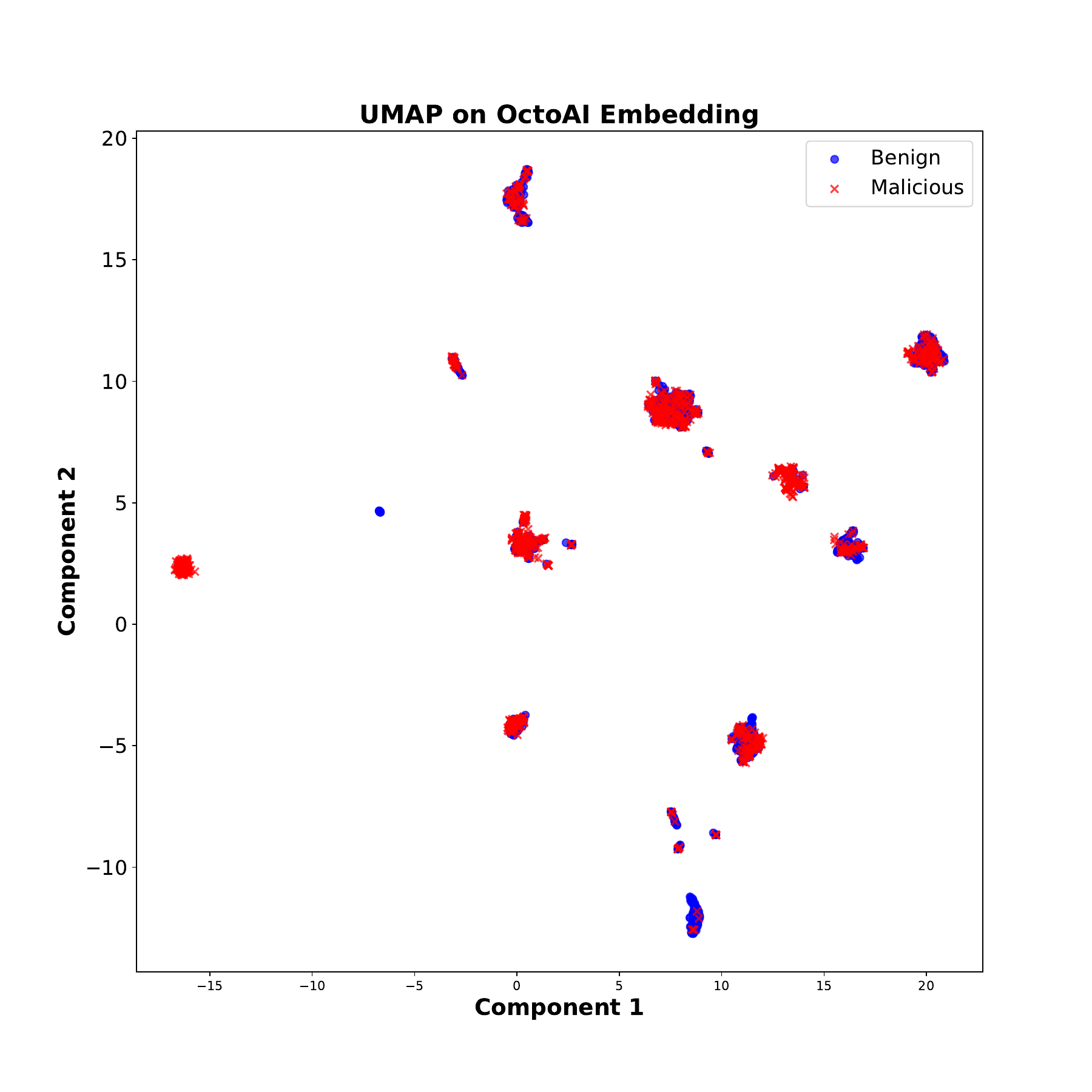}
    \includegraphics[width=.33\textwidth]{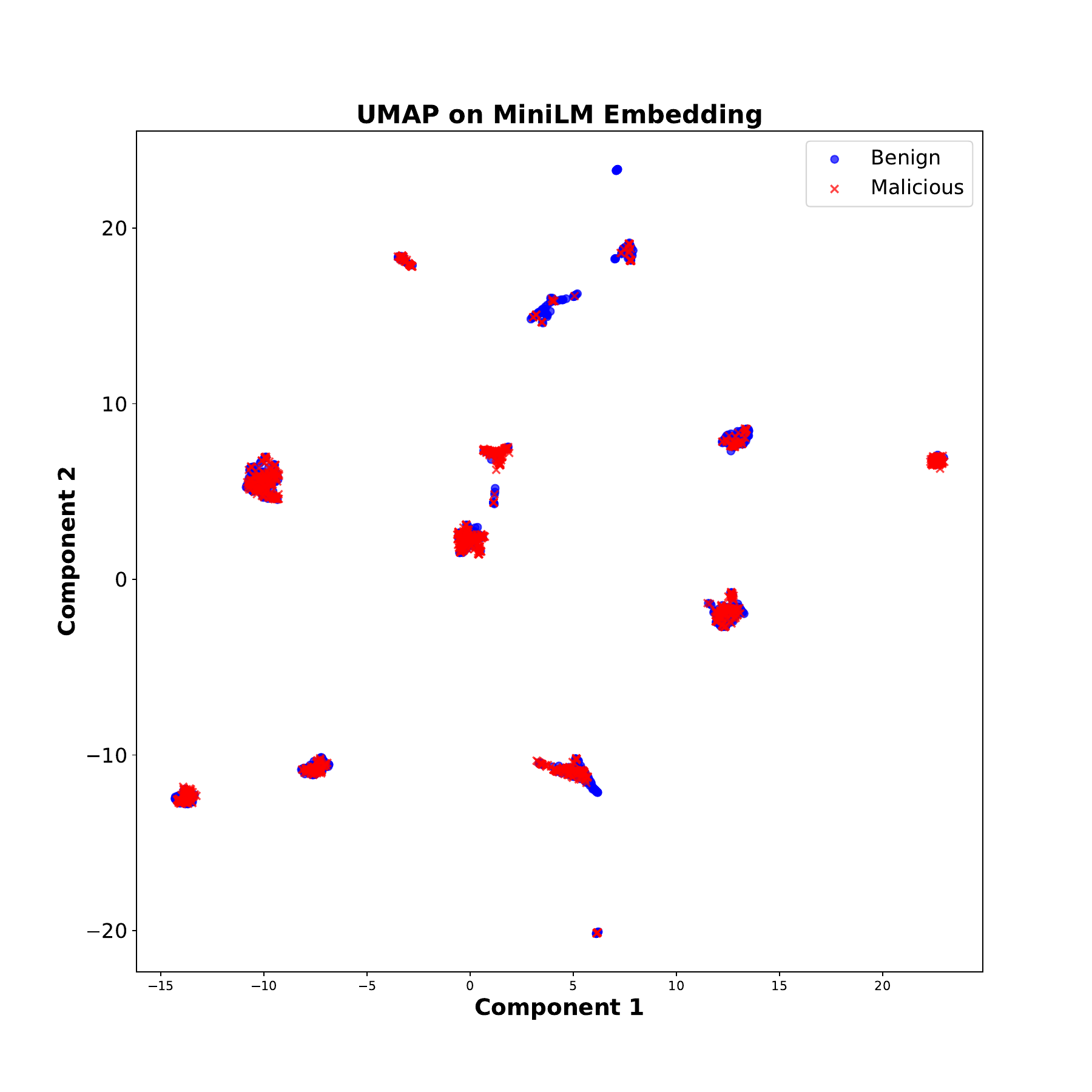}
    \caption{Visualization of OpenAI (Left), GTE (Middle), and MiniLM (Right) embedding distribution after applying Uniform Manifold Approximation and Projection (UMAP).}
    \label{fig:umap_on_embedding}
\end{figure*}

The first phase of our experiments is centered around visualizing the low-dimensional projections of the embeddings, generated using PCA, t-SNE, and UMAP. With details provided in Table \ref{tab:captured_information}, we capture 12.62\% to 15.83\% and 10.83\% to 11.43\% of the information for all three embeddings through PCA with the 1st and 2nd Principal Components, respectively. We present the visualizations of OpenAI, OctoAI, and MiniLM embeddings after applying PCA in Fig. \ref{fig:pca_on_embedding}. For t-SNE, our experiments involve investigating visualizations using perplexity values ranging from 5 to 50. We achieve the most well-separated clusters by selecting a perplexity of 15 (Fig. \ref{fig:tsne_on_embedding}). Lastly, we depict the visualizations of all three embeddings after applying UMAP in Fig. \ref{fig:umap_on_embedding}.

\begin{table*}[t]
    \centering
    \caption{Details of the captured information after reducing the dimensionality of datasets.}
    \begin{tabular}{ c c c c }
        \hline
        Principal Component & OpenAI & GTE & MiniLM \\ \hline
        1st Principal Component & 13.59\% & 15.84\% & 12.62\% \\
        2nd Principal Component & 10.83\% & 11.03\% & 11.43\% \\\hline
    \end{tabular}
    \label{tab:captured_information}
\end{table*}

As seen in the three plots, we do not find clear separations between benign and malicious data points. Especially, linear or sigmoid separations are not observed between red and blue clusters. This indicates that tree-based and/or gradient boosting algorithms will be better suited to separate the malicious data points compared to methods based on the linearity assumption, such as logistic regression and linear discriminant analysis.

\subsection{Answer to RQ2: Can we effectively identify malicious prompts to thwart prompt injection attacks?}
As mentioned in the earlier section, we employ Logistic Regression, XGBoost, and Random Forest classifiers for performing binary classification tasks. We maintain the size of all embeddings consistently for both training and testing purposes. Initially, we report their performances in terms of AUC (Area under the ROC Curve). The AUC value ranges between 0 and 1 ---the higher the value, the better a classifier's prediction capability. For example, a classifier with perfect predictions would have an AUC of 1. Table \ref{tab:auc} presents the AUC values for all classifiers across different embeddings. We observe that Random Forest consistently outperforms the other two classifiers under all experimental settings. Among the embeddings methods, OpenAI performs the best---possibly owing to its higher dimensionality. 

\begin{table*}[t]
    \centering
    \caption{Performance comparisons based on AUC across classifiers and embedding methods.}
    \begin{tabular}{c c c c }
        \hline
        Embedding & Logistic Regression & XGBoost & Random Forest \\ \hline
        OpenAI & 0.637 & 0.726 & 0.764 \\
        GTE & 0.612 & 0.690 & 0.731 \\
        MiniLM & 0.608 & 0.687 & 0.730 \\\hline
    \end{tabular}
    \label{tab:auc}
\end{table*}

Using default binary predictions from the ML classifiers, we also compute precision, recall, and F1 scores, which also range between 0 and 1.

\[ \text{Precision} =  \frac{\text{True Positive}}{\text{True Positive + False Positive}}\]

\[ \text{Recall} =  \frac{\text{True Positive}}{\text{True Positive + False Negative}}\]

$$ \text{F1} = 2 \times \frac{\text{Precision} \times \text{Recall}}{\text{Precision} + \text{Recall}}$$

Similar to AUC, Random Forest performs the best compared to XGBoost and Logistic Regression, as highlighted in Table \ref{tab:precision-recall}. For instance, the precision and recall scores of Random Forest for OpenAI embeddings are 86.65\% and 86.96\%, respectively, which are up to 3\% higher than XGBoost and 6\% higher than Logistic Regression. Similar results are observed for the other two embeddings as well. Therefore, the Random Forest algorithm is identified as the best classifier in our study.

\begin{table*}[t]
    \centering
    \caption{Binary classification performance of embedding-based ML classifiers.}
    \begin{tabular}{ c c c c c c c c c c }
        \hline
        & \multicolumn{3}{c }{Logistic Regression} & \multicolumn{3}{c }{XGBoost} & \multicolumn{3}{c }{Random Forest} \\ 
        Embedding & Precision & Recall & F1 & Precision & Recall & F1 & Precision & Recall & F1\\ \hline
        OpenAI & 0.793 & 0.807 & 0.80 & 0.832 & 0.841 & 0.836 & 0.867 & 0.867 & 0.867 \\
        GTE & 0.785 & 0.799 & 0.792 & 0.820 & 0.830 & 0.825 & 0.849 & 0.853 & 0.851\\
        MiniLM & 0.777 & 0.795 & 0.789 & 0.820 & 0.829 & 0.824 & 0.849 & 0.853 & 0.851\\\hline
    \end{tabular}
    \label{tab:precision-recall}
\end{table*}

\paragraph{Comparison with State-of-the-Art Classifiers.} We compare the performances of our embedding-based classifiers with four state-of-the-art deep learning classifiers available on Hugging Face. To begin describing each classifier, \citet{tunstall2022efficient} released their Sentence Transformer model on Hugging Face as \textit{Myadav: setfit-prompt-injection-MiniLM-L3-v2}\footnote{\url{https://huggingface.co/Myadav/setfit-prompt-injection-MiniLM-L3-v2}} for text classification. The scond classifier we compared with is \textit{protectai: deberta-v3-base-prompt-injection}, which is based on DeBERTaV3 \cite{he2020deberta} and was released in 2023\footnote{\url{https://huggingface.co/protectai/deberta-v3-base-prompt-injection}}. We also compare against \textit{protectai: deberta-v3-base-prompt-injection-v2}, an updated version of the above model based on optimization of hyperparameters, training regimens, and dataset compositions\footnote{\url{https://huggingface.co/protectai/deberta-v3-base-prompt-injection-v2}}. Finally, we compare against another popular finetune of DeBERTaV3 called \textit{deepset: deberta-v3-base-injection}\footnote{\url{https://huggingface.co/deepset/deberta-v3-base-injection}}. We examine the performances of all four classifiers using AUC, precision, and recall scores on our test dataset including both malicious and benign prompts.

\begin{table*}[t]
    \centering
    \caption{Performance comparisons against popular and accurate prompt injection classifiers available on Hugging Face.}
    \begin{tabular}{ l c c c c }
        \hline
        Model & AUC & Precision & Recall & F1 \\ \hline
        Myadav: setfit-prompt-injection-MiniLM-L3-v2 & 0.594 & 0.827 & 0.62 & 0.709\\
        protectai: deberta-v3-base-prompt-injection & 0.531 & 0.774 & 0.910 & 0.837\\ 
        protectai: deberta-v3-base-prompt-injection-v2 & 0.511 & 0.758 & \textbf{0.991} & 0.859\\
        deepset: deberta-v3-base-injection & 0.500 & 0.762 & 0.988 & 0.860\\
        Our best (Random Forest + OpenAI) & \textbf{0.764} & \textbf{0.867} & 0.870 & \textbf{0.868}\\ \hline        
    \end{tabular}
    \label{tab:performance-comparison}
\end{table*}

Table \ref{tab:performance-comparison} reports the performance metrics for the four SoTA models and compares them against our best model. The Random Forest classifier built with OpenAI embeddings outperforms other SoTA classifiers in terms of AUC and precision scores. The recall score for our best model is 86.96\%, whereas other classifiers achieve scores in the high 90s. However, due to higher precision, our classifier achieves the highest F1 score. In real-world ML pipelines, calibrating and updating thresholds based on test time data is standard practice, and balancing precision and recall is important. In general, the three DeBERTa finetunes have high recall but low precision, whereas \textit{Myadav: setfit-prompt-injection-MiniLM-L3-v2} has high precision but low recall. Our embedding-based approach strikes a better balance between these two metrics.

\section{Discussion and Conclusion}
In this paper, we propose a novel embedding-based classifier approach to detect malicious prompts that lead to successful prompt injection. We curate a large dataset of benign and malicious prompts from several repositories on Hugging Face and generate their embeddings using three methods. Through two-dimensional visualizations, we investigate the distributional differences of embeddings labeled benign and malicious and perform binary classification tasks using a number of supervised ML classifiers. The Random Forest classifier trained using OpenAI embeddings exhibits the best performance, achieving an AUC of 0.764, precision of 0.867, and recall of 0.87. Comparing our classifier's performance with several SoTA prompt injection available on Hugging Face designed for similar tasks. Our results demonstrate that our classifier surpasses all of them in terms of AUC and precision scores.

One of the main goals of this paper was to investigate the dissimilarities between embeddings of malicious and benign prompts using dimensionality reduction algorithms. However, we did not find a clear linear separation in the generated visualizations. While we leave further exploration for future work, we did achieve commendable performance in detecting prompt injections via traditional ML classifiers trained on these embeddings. Especially, the random forest classifier was able to outperform the most popular and highest performant models available in the open source.

Our study examines the efficacy of traditional ML classifiers. Neural network-based classifiers may also be constructed based on embeddings. This needs to be explored in future work. Our research has primarily focused on crafting classifiers to detect direct prompt injections. A similar approach can also be taken to craft embedding-based detectors for other LLM attack vectors and failure modes, which are indirect prompt injections, toxicity, and hallucination.

\begin{acknowledgments}
We thank the CAMLIS program committee and reviewers for reviewing the paper and sharing valuable feedback that led to significant improvements.
\end{acknowledgments}

\bibliography{sample-ceur}


\end{document}